\documentclass
[showpacs,prl,amsfonts,amssymb,oneside,twocolumn,notitlepage,10pt,preprint,balancelastpage,a4paper]{revtex4}%
\usepackage{amsfonts}
\usepackage{amsmath}
\usepackage{amssymb}
\usepackage{graphicx}%
\begin{document}
\title{Controled observation of a nonequilibrium Ising--Bloch transition in a
nonlinear optical cavity}
\author{A. Esteban--Mart\'{\i}n, V. B. Taranenko, J. Garc\'{\i}a, G. J. de
Valc\'{a}rcel, and Eugenio Rold\'{a}n}
\affiliation{Departament d'\`{O}ptica, Universitat de Val\`{e}ncia, Dr. Moliner 50,
46100--Burjassot, Spain}

\begin{abstract}
We report the controled observation of the nonequilibrium Ising--Bloch
transition in a broad area nonlinear optical cavity (a quasi--1D single
longitudinal--mode photorefractive oscilator in a degenerate four--wave mixing
configuration). Our experimental technique allows for the controlled injection
of the domain walls. We use cavity detuning as control parameter and find that
both Ising and Bloch walls can exist for the same detuning values within a
certain interval of detunings, i.e., the Ising--Bloch transition is hysteretic
in our case. A complex Ginzburg--Landau model is used for supporting the observations.

\end{abstract}

\maketitle

Spatially extended bistable systems exhibit a large variety of localized
structures as bistability allows that different states ocupy different spatial
regions. An interesting subclass is that of systems with broken phase
invariance \cite{Staliunas}, which can display defects in the form of
interfaces, so-called fronts, across which the system passes from one of the
bistable phases to the other one in adjacent spatial domains. Here we consider
one--dimensional systems.

In equilibrium systems, characterized by a well defined free energy, two cases
can be distinguished depending on whether the two states that the front
connects have equal or different free energy: When equivalent (e.g., the two
possible orientations of magnetization in a ferromagnet) fronts are static,
whilst when both states have different free energy, fronts move so that the
lower energy state finally invades the whole system. Front dynamics is richer
in nonequilibrium systems whose dynamics does not derive from a free energy.
As in equilibrium systems, a front connecting two non--equivalent states is a
transient state. Contrarily, the behaviour is different from equilibrium
systems when the front connects two equivalent states (these fronts are known
as domain walls, DWs): Motion in this case is still possible through a parity
breaking bifurcation occurring at the front core \cite{Coullet}. Before the
bifurcation the (resting) structure is \textquotedblleft odd\textquotedblright%
\ with respect to the front and is known as Ising wall; after the bifurcation
that symmetry is lost and the structure, known as Bloch wall, moves. This
nonequilibrium Ising--Bloch transition (NIBT) \cite{Michaelis} ---borrowing
its name from the (equilibrium) Ising--Bloch transition of ferromagnets
\cite{Bualevskii}--- is generic in self-oscillatory systems parametrically
forced at twice their natural oscillation frequency
\cite{Coullet,deValcarcel02}.

There are few experimental observations of this phenomenon. As far as we know,
it has been reported only in liquid cristals \cite{Meron} either subjected to
rotating magnetic fields, \cite{Frisch,Nasuno} or to an alternate electrical
voltage \cite{Kawagishi}. This last experiment constitutes a particularly
clear observation of the NIBT free of 2D effects, which complicate front
dynamics through curvature effects.

Nonlinear optical cavities provide several examples of systems with broken
phase invariance for which the NIBT has been predicted but not yet observed
(intracavity type II second--harmonic generation \cite{Michaelis}, degenerate
optical parametric oscillation \cite{Izus00,Skryabin01,Perez04}, and cavity
degenerate four--wave mixing \cite{Sanchez04}). In an optical system, the DW
connects two stable states of equal light intensity but opposite (i.e.
separated by $\pi$) phase. In the Ising wall the phase jumps sharply by $\pi$,
and the light intensity is null at the wall core, whilst in the Bloch wall the
phase angle rotates continuously through $\pi$, and the light intensity is
minimal but not null at the wall core. As two senses of rotation are possible,
Bloch walls are chiral (have the symmetry of a corkscrew), what has important
dynamical consequences as walls with opposite chirality move in opposite
directions \cite{Coullet}. In this Letter we report the first experimental
observation of the NIBT in a nonlinear optical system. Moreover, the NIBT we
describe below presents a distinctive feature that, to the best of our
knowledge, has not been described before: It is a hysteretic NIBT, i.e., there
is a domain of coexistence of both Ising and Bloch walls.

Our system is a single longitudinal--mode photorefractive oscillator in a
degenerate four-wave mixing configuration
\cite{Taranenko98,Larionova04,Esteban04}, exactly the same system used in
\cite{Esteban04}. Two counterpropagating coherent pump beams of equal
frequency illuminate a \textrm{BaTiO}$_{3}$ crystal which produces scattered
light, part of which seeds the oscillation into the (plane--mirrors)
Fabry-Perot resonator. Oscillation is dynamically ruled by gain, losses,
cavity detuning (the mismatch between the frequency of the pump fields and
that of the longitudinal cavity mode in which the system oscillates), and
diffraction. Each of these parameters can be modified in our experiment up to
a certain extent. In particular, the detuning, $\Delta$, can be finely tuned
by our adjusting of the stabilized cavity length \cite{Esteban04} and plays
the role of control parameter in our experiment. Two characteristics of the
cavity are particularly relevant. First, we are using a near self-imaging
resonator \cite{Arnaud69} what allows to have a very large Fresnel number,
i.e. allows the oscillation of a very large number of transverse modes.
Second, we intentionally make our system quasi--1D in the transverse dimension
by placing slits in the Fourier planes. The width of the slits is adjusted to
the size of the diffraction spot in these planes. In this way, beams with an
inclination such that their transverse wavevector does not lie in the plane
defined by the center line of the slit are not compatible with the diffraction
constrains of the cavity. The Fourier filtering allows the use of finite width
slits and still get rid of most 2D effects, such as front curvature. All this
amounts to saying that our nonlinear cavity is equivalent to a large aspect
ratio 1D nonlinear system. Figure 1 displays a simplified scheme of the
photorefractive oscillator and the reader is referred to \cite{Esteban04} for
full details.

In \cite{Larionova04} the dynamics of spontaneously formed phase domains was
experimentally studied in a setup very similar to the one we use here but,
importantly, two--dimensional in the transverse plane. Larionova \textit{et
al. }\cite{Larionova04} characterize the closed wall separating different
phase domains and find Bloch--type and Ising--type segments along the same
wall. They did not observe, however, any NIBT because of the 2D character of
their system.

In \cite{Esteban04} pattern formation in the same system we are using here was
discussed. The basic bifurcation diagram found in \cite{Esteban04} can be
summarized as follows: For large enough negative detuning, $\Delta$, the
system forms a periodic structure (stripe pattern) whose spatial frequency
varies as $k^{2}\sim-\Delta$ (this law allows the sharp determination of the
cavity resonance); for small negative $\Delta$, aperiodic patterns are formed;
and for positive $\Delta$, homogeneous states are found. In this last region,
DWs (isolated or not) form spontaneously, as already described in
\cite{Esteban04}. In this letter we address the dynamic behaviour of these DWs
as $\Delta$ is varied.

A crucial difference with respect to \cite{Esteban04} and from any previous
experiment is that in the present experiment DWs are created in a controled
way. This is achieved by injecting, for a short time, a tilted laser beam of
the same frequency as the pumps into the photorefractive oscillator. This
tilted beam has a phase profile $\phi\left(  x\right)  =\phi_{0}+\left(
2\pi/\lambda\sin\alpha\right)  x$ where $\phi_{0}$ is a constant, $\lambda$ is
the light wavelength, $\alpha$ is the tilt angle with respect to the resonator
axis (that can be varied in the experiment) and $x$ is the spatial coordinate.
As the degenerate four-wave mixing process is phase sensitive \cite{Yeh}, only
two phase values (say $0$ and $\pi$, modulo $2\pi$) are amplified and, as a
consequence, the portions of the transverse plane whose phase lies in $\left]
-\pi/2,+\pi/2\right[  $ are attracted towards the phase value $0$ and those
that lie in $\left]  +\pi/2,+3\pi/2\right[  $ are attracted towards the phase
value $\pi$. The points at which the phase is exactly $\pm\pi/2$ (modulo
$2\pi$) are dark as those phase values are not amplified and, more
importantly, because these points are separating adjacent domains with
opposite phases. These points (the domain walls) are thus topological defects
\cite{Trillo97}. Once DWs are envisaged, the injection is removed. In this way
a single DW (or more, for larger tilt angles in the writing beam) can be
written at the desired location along the transverse dimension.

The intensity distribution on the $x-y$ plane, perpendicular to the resonator
axis $z$, is not homogeneous due to several reasons, the gaussian profile of
the pump beam among them. On the other hand, the intracavity slits make the
system quasi--1D but, obviously, not strictly 1D. Then in order to analyze
quantitatively the intensity and phase properties of the field, either a
particular value of $y$ is chosen or some averaging in the $y$ direction is
made. Our strategy has consisted in isolating the central region in the $y$
direction where the output intensity is approximately constant (for fixed $x$)
and then making an average of the field inside this region. We chose this
option because this procedure gets rid of local imperfections, which could
affect the wall characterization if a particular $y$ value were chosen. Notice
that in this way the outer borders in the $y$ direction are discarded, a
convenient procedure in order to avoid border effects in the DW
characterization.The interferometric reconstruction technique giving the
complex field $A\left(  x,y\right)  $ is described in full detail in
\cite{Larionova04} and it is from these data that we obtain the average field
$\left\langle A\left(  x\right)  \right\rangle _{y}$.

Depending on the cavity detuning value, two different types of DWs are found:
For small (large) values of $\Delta$, static Ising type (moving Bloch type)
DWs are seen. The Ising or Bloch character of the DWs is clearly appreciated
in Fig. 2: The Ising wall exhibits a discontinuous phase jump and the field
intensity is nearly zero at the wall center (Fig. 2a), whilst in the Bloch
wall the phase variation is smooth and the field intensity minimum is clearly
different from zero (Fig. 2b). But the most striking difference between Ising
and Bloch walls is in their different dynamic behaviour. In Fig. 3
interferometric snapshots of Ising (Fig. 3a) and Bloch (Fig. 3b and 3c) DWs
are shown. The difference between Figs. 3b and 3c lies in the sign of the
velocity, which is different due to the different chirality of the Bloch walls
in the two figures. It can be appreciated that the wall velocity is not
constant, i.e., there is some acceleration. This is due to the spatial
inhomogeneity of the field along the $x$--direction as the intensity gradient
introduces spurious effects on the wall dynamics. Then, when evaluating the
wall velocity (see below), we discard the part of the trajectory where the
acceleration is more obvious.

Let us see now where (in terms of the cavity detuning $\Delta$, the control
parameter) and how the NIBT occurs. In order to clarify this we proceded as
follows: Starting with one of the Ising walls, $\Delta$ is increased in small
steps \cite{steps}. We observe that the Ising wall remains at rest until some
critical detuning value is reached, call it $\Delta_{\mathrm{IB}}$, where the
wall spontaneously starts to move (Fig. 3b or 3c), thus signalling a NIBT. As
the illuminated region is finite, after some time the Bloch wall dissapears
through the illuminated border. Then, in order to follow the dynamics for
$\Delta>\Delta_{\mathrm{IB}}$, new DWs are injected for each increasing value
of $\Delta$. We check in this way that for $\Delta>\Delta_{\mathrm{IB}}$ walls
are always of the Bloch type. Next we proceed to a reverse scan of the
detuning: Starting with large positive $\Delta$, we inject a DW and see that
it is of Bloch type. Then we decrease $\Delta$ and repeat the operation until
a value of $\Delta$ is reached, say $\Delta_{\mathrm{BI}}$, below which only
Ising walls are observed. The interesting feature is that $\Delta
_{\mathrm{BI}}<\Delta_{\mathrm{IB}}$, i.e., there is a domain of detuning
values ($\Delta_{\mathrm{BI}}<\Delta<\Delta_{\mathrm{IB}}$) were both Ising
and Bloch walls exist. In fact we have checked repeatedly that within this
domain of hysteresis both Bloch and Ising walls are observed alternatively
after subsequent wall injections for fixed $\Delta$ (in this detuning range
the final state obviously depends on the initial condition, which seemingly we
are not able to control). We repeated the procedure until we convinced
ourselves of the reproducibility of the observation. Let us remark that the
hysteresis cannot be attributed to mechanical hysteresis in the piezoelectric
mirror that controls $\Delta$, as subsequent observations of the two types of
walls \textit{for fixed} $\Delta$ rule out this possibility. Figure 4
summarizes these findings. The upper and lower arrows indicate the values of
$\Delta$ where the NIBTs are observed (which we have called $\Delta
_{\mathrm{BI}}$ and $\Delta_{\mathrm{IB}}$), and the squares (crosses) denote
increasing (decreasing) detuning scan. Notice that the smooth decrease of the
velocity in the upper branch finds a sharp end at a detuning of 5\%
\textit{ca} of the cavity free spectral range (FSR). These results firmly
stablish the existence of the hysteresis cycle.

We pass to interpret the observation: Although one could think that the
observed hysteresis could be due to a subcritical character of the NIBT in our
system, we argue that it is due to the existence of a hysteresis cycle in the
homogeneous states of the system. In the first case, the homogeneous states
connected by the DW, differing only in their phase, would have the same
intensity irrespectively of the wall type (Ising or Bloch). Contrarily, in the
second case the homogeneous states should be different for Ising and Bloch
walls. The latter is our case: The two equivalent homogeneous states (of
opposite phases) connected by the Ising wall have larger intensities than
those connected by the Bloch wall. This implies that there are two homogeneous
states in the system that are bistable in a limited range of cavity detuning
values. We have measured the homogeneous state intensity by scanning $\Delta$
and have found evidence of the existence of a bistability loop in the
homogeneous state. Very likely, the NIBT occurs somewhere in the middle of the
unstable branch of the homogeneous state (the one connecting the higher and
the smaller intensity stable states). This would explain why we never observe
Bloch walls moving at small velocities.

Finally, in order to provide some theoretical support we consider the
following phenomenological, adimensional model%
\begin{align}
\partial_{t}E  &  =\left(  1-i\Delta\right)  E+\gamma E^{\ast}+\left(
1+i\alpha\right)  \partial_{x}^{2}E\nonumber\\
&  -\left(  1+i\beta_{3}\right)  \left\vert E\right\vert ^{2}E-i\beta
_{5}\left\vert E\right\vert ^{4}E, \label{equ}%
\end{align}
where $E$ is proportional to the complex intracavity field amplitude, $\gamma$
accounts for parametric gain (the typical phase symmetry breaking term of
degenerate four-wave mixing \cite{Sanchez}), $\Delta$ is the detuning,
$\alpha$ accounts for diffraction, and $\beta_{3}$ and $\beta_{5}$ account for
nonlinear dispersion. Without the fifth order nonlinearity, Eq. (\ref{equ}) is
the complex Ginzburg--Landau equation investigated in \cite{Coullet}. This
minimum correction to the model of \cite{Coullet}, the term $-i\beta
_{5}\left\vert E\right\vert ^{4}E$, appears after Taylor expanding the
photorefractive nonlinearity that has a saturating form. We have found that
Eq. (\ref{equ}) displays an intensity-bistable homogeneous state in
coincidence with a hysteretic NIBT (similar to Fig. 4) for a wide range of
parameters (as an example, for $\gamma=2$, $\alpha=1$, $\beta_{3}=-2.7$, and
$\beta_{5}=1$, the hysteretic NIBT is observed; details will be given
elsewhere). Of course, we do not claim that this constitutes a theoretical
interpretation of our experimental results. Nevertheless this numerical
finding supports clearly our observations as Eq. (\ref{equ}) contains the
basic physical phenomena present in the experimental system.

In conclusion, we have reported the first observation of the nonequilibrium
Ising--Bloch transition in an optical system, a quasi--1D single
longitudinal--mode photorefractive oscilator in a degenerate four--wave mixing
configuration, using the cavity detuning as the control parameter. The NIBT we
report is special in the sense that it is hysteretic, i.e., in terms of the
control parameter, there is a domain of coexistence of both Ising and Bloch
walls. The origin of this hysteretic cicle lies in the bistability exhibited
by the homogeneous state of the system.

\begin{acknowledgments}
This work has been financially supported by Spanish Ministerio de Ciencia y
Tecnolog\'{\i}a and European Union FEDER (Projects BFM2002-04369-C04-01 and
BFM2001-3004), and by Ag\`{e}ncia Valenciana de Ci\`{e}ncia i Tecnologia
(Project GRUPO S03/117). V.B.T. is financially supported by the Spanish
Ministerio de Educaci\'{o}n, Cultura y Deporte (grant SAB2002-0240). We
gratefully acknowledge fruitful discussions with C.O. Weiss and continued
support by C. Ferreira.
\end{acknowledgments}

\bigskip

{\LARGE Figure Captions}

\textbf{Fig. 1.-} Scheme of the experimental setup. M: cavity mirrors; L:
effective cavity length; D: diaphragm that makes the system quasi-1D; PZT:
piezo--transducer for control of the cavity detuning\thinspace; and \textit{f}
: lenses focal length. Photographs: output field intensity in the near and far
fields for a striped pattern.

\textbf{Fig. 2.-} DWs amplitude (dashed line, arbitray units ) and phase (full
line) profiles as reconstructed with the interferometric technique in
\cite{Larionova04}. They correspond to the cases represented in the fourth row
snapshots in Figs. 3 (a) and (b), respectively.

\textbf{Fig. 3.-} Interferometric snapshots of Ising and Bloch walls. In Fig.
3a (3b and 3c) the wall is of Ising (Bloch) type, and the detuning is 0 MHz
(25 MHz). Time runs from top to bottom in steps of 5 s. The transverse
dimension is 1.6 mm.

\textbf{Fig. 4.-} DW velocity versus cavity detuning (in units of the cavity
free--spectral range) showing the hysteretic cycle. The two NIBTs are marked
with arrows. The cavity free spectral range is 120 MHz.


\begin{thebibliography}{99}
\bibitem {steps}The detuning was varied, tipically, in steps of 1Volt in the
piezoelectric mirror that controls the cavity length. Sometimes smaller steps
of 0.5Volts were used. A complete free spectral range is scanned with 33Volts.
\end{thebibliography}
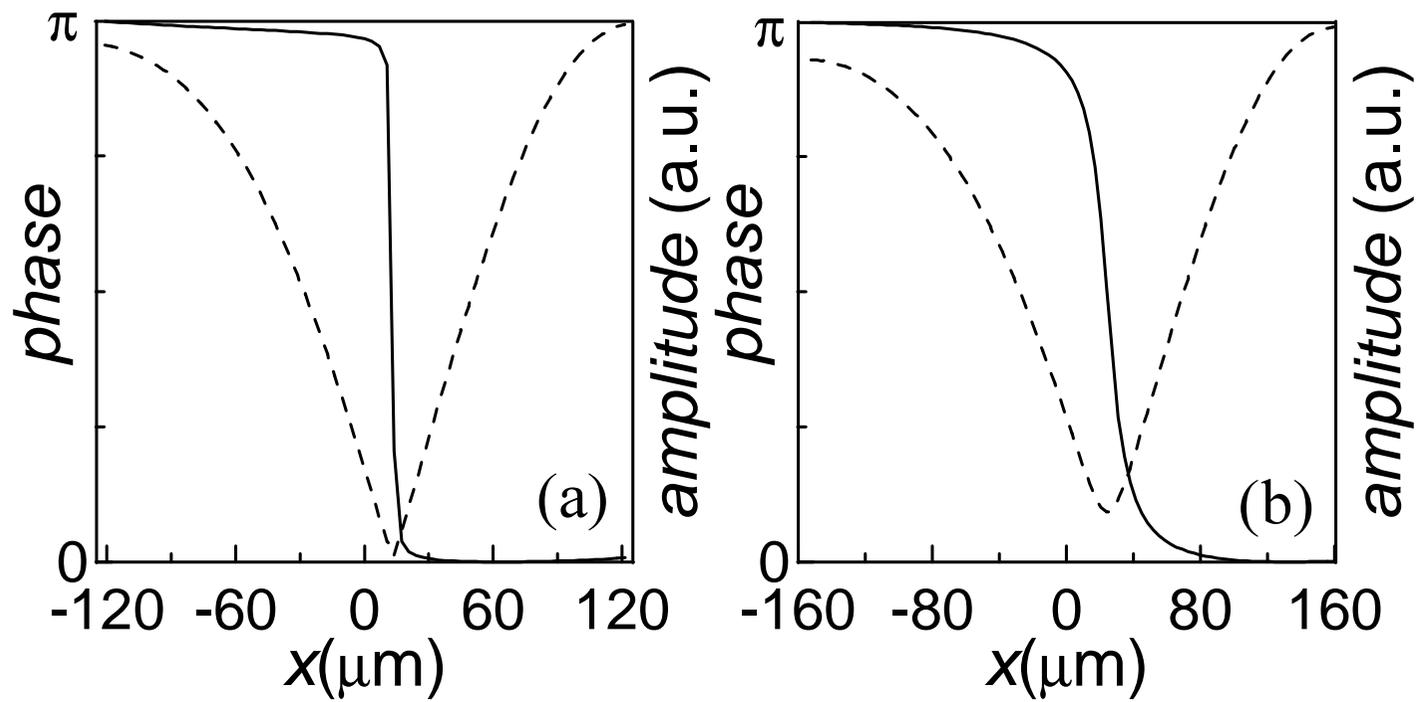
\end{document}